# Photo-Hall effect spectroscopy with enhanced illumination in p-Cd$_{1-x}$Mn$_x$Te showing negative differential photoconductivity

A. Musiienko[1,a)], R. Grill[1], P. Moravec[1], P. Fochuk[2], I. Vasylchenko[1], H. Elhadidy[3,4] and L. Šedivý[1]

[1]*Institute of Physics, Charles University, Prague, 121 16, Czech Republic*

[2]*Chernivtsi National University, Ukraine*

[3]*Central European Institute of Technology, Institute of Physics of Materials, ASCR, Brno 61662, Czech Republic*

[4]*Faculty of Science, Physics Department, Mansoura University, Mansoura 35516, Egypt*

Abstract:

We studied deep levels (DLs) in p-type Cd$_{1-x}$Mn$_x$Te by photo-Hall effect spectroscopy with enhanced illumination. We showed that the mobility of minority (925±11 cm$^2$s$^{-1}$V$^{-1}$) and majority (59.6±0.4 cm$^2$s$^{-1}$V$^{-1}$) carriers can be deduced directly from the spectra by using proper wavelength and excitation intensity. Four deep levels with ionization energies $E_{t1} = E_V + 0.63$ eV, eV, $E_{t2} = E_V + 0.9$ eV, $E_{t3} = E_C - 1.0$ eV and $E_{t4} = E_C - 1.3$ eV were detected and their positions in the bandgap were verified by comparison of photogenerated electron and hole concentrations. Deduced DL model was analyzed by numerical simulations with Shockley-Reed-Hall charge generation-recombination theory and compared with alternative DL models differing in the position of selected DLs relative to $E_c$ and $E_v$. We showed that the consistent explanation of collected experimental data principally limits the applicability of alternative DLs models. We also demonstrated the importance of the extended operation photon fluxes (I > 4×10$^{14}$ cm$^{-2}$s$^{-1}$) used in the spectra acquisition for correct determination of DLs character. Negative differential photoconductivity was observed and studied by charge dynamic theoretical simulations.

---

[a] E- mail: musienko.art@gmail.com

# 1 Introduction

Inevitable crystallographic defects such as point, line, planar and bulk defects represent one of the most serious problems currently faced by a semiconductor manufacture technology. The presence of such defects in the semiconductor lattice leads to the formation of energy states in the band gap. These states can be divided into two major groups. The first one, shallow levels, are the less localized states with activation energies $E_t$ typically less or similar to the thermal energy kT. Shallow states are produced by small lattice perturbations and generally act as dopants (donors or acceptors) without harmful influence on the material. The second group is called deep levels (DLs) with $E_t$ significantly exceeding kT. The difference from the previous case of shallow states consists in much larger perturbation brought by these defects to the lattice potential in the close vicinity of the defect. Thus, the wave function of the trapped electron (or hole) is much more localized. Deep levels have a harmful impact on the utilization of the semiconductor in optoelectronic applications acting as electron (hole) traps or recombination centers depending on their cross sections ratio $\sigma_e/\sigma_h$, where $\sigma_e$ ($\sigma_h$) is the electron (hole) capture cross-section. Many different deep level spectroscopy methods [1–4] can identify deep-level (DL) ionization energy but a common problem is to identify relative DL position inside the band gap, i.e. whether DL energy locates below or above the Fermi energy. The knowledge of the exact position of DL in the bandgap and its trapping characteristics plays a crucial role in the semiconductor characterization [5]. Multiple achievements were attained by photo-Hall effect spectroscopy (PHES). It was used to prove the bipolar character of photo-conduction in high resistivity p-CdTe:Cl [6]. Low-temperature (10–40 K) PHES was successfully used for the characterization of impurities in p-HgCdTe [7]. A phenomenological model of deep donors in GaSb:S was developed and supported by PHES [8,9]. The character of DLs was revealed in semi-insulating GaAs, where all identified traps were found to be electron-type [10]. Another approach to measure the mobility of photo-carriers by polarization-sensitive terahertz emission in the magnetic field was presented by Lin et al. [11]. It was found by Chen et al. [12,13] that PHES can be used for characterizing hybrid organic-inorganic lead $CH_3NH_3PbCl_3$($Br_3$ or $I_3$) perovskites. PHES measurements in materials with low carrier mobility or suppressed Hall signal [14] can be used with a combination of the alternating magnetic field and lock-in amplifier [13]. In such a set-up the magnetic signal is chosen as the reference signal. The majority and minority carrier mobility are important material parameters for many semiconductor devices such as solar cells, radiation detectors, diodes, transistors, etc. While the majority carrier mobility can be obtained by several techniques [5,15,16], the minority carrier mobility is hardly accessible.

The PHES measurements can be divided into two groups: the Hall signal measurements as a function of illumination intensity at various temperatures and the Hall signal measurements with different illuminating photon energies. Some modifications of the experiment are possible inside the groups such as dual photon energy illumination [17] or time transients measurements [18]. In our previous research [18], we pointed out that PHES faces a serious problem in the p-type material where strong recombination of minority carriers (electrons) suppresses electron impact in Hall mobility ($\mu_H$) and photoconductivity (PhC). We also showed that the determination of DLs position in the band gap for n-type CdZnTe is complicated and can be accessed in dual-wavelength illumination regime [17]. Because of these limitations, scant interest was paid to PHES by researchers and the benefits of the method were underrated.

Negative differential photoconductivity (NDPC) is often observed in optical transport measurements of semiconductors. It is characterized by a decrease of the photoconductivity under enhanced illumination [19]. The effect of the conductivity decrease under illumination below the dark equilibrium value is known as absolute negative photoconductivity (ANPC) [20]. According to known models, NDPC and ANPC can be induced by two mechanisms. The first one is associated with a decrease of the carrier mobility induced by light radiation. This effect can be reached by stimulation of scattering mechanisms as it was shown for graphene [11,21], gold nanoparticles [22], Ge(Si) type-II quantum dots [23], and interacting quantum gas [24]. NDPC stimulated by potential non-uniformity was observed by our group in CdZnTe [Musiienko JAP] and by Shalish et al. [25] in GaAs films using Hall photovoltage. The second mechanism producing NDPC is the decrease of the majority carrier concentration due to recombination of minority carriers supported by a recombination-type crystallographic defect. This model was first proposed by Stockman et al. [26] and implemented for diamond photodetectors [20], thin films [27], $MoS_2$ monolayers [28], p-type PbEuTe and other materials. A similar mechanism of enhanced radioactive capture of hot electrons was proposed by Ridley et al. [29] for III-V and II-VI semiconductors. Despite a large number of experimental observations, the detailed theoretical study of NDPC (or ANPC) induced by defect levels was not presented. A simple model was given for inhomogeneous materials by Bube et al [30]. Despite the similar nature of the two effects the connection between NDPC and ANPC was not shown. The dependence of NDPC on DL parameters was also not specified.

The binary compound semiconductor CdTe and its ternary modifications $Cd_{1-x}Zn_xTe$ and $Cd_{1-x}Mn_xTe$ have a lot of applications in X- and gamma-ray detection [31,32] and in photovoltaics [33] due to a convenient tunable band gap and good absorption properties [34]. Despite these benefits, as grown materials often suffer from crystallographic defects that deteriorate the transport properties of the devices [35]. The CdTe-based material resistivity homogeneity and detection properties can be enhanced by adding Se or Mn into the solution [36,37] due to uniform segregation of these components. Moreover, the bandgap energy can be controlled by the compound composition [38]. Such compound materials have known limitations for detector applications such as strong carrier recombination [39] and device polarization [40,41] caused by deep levels inside the band gap.

In this study, we combine the PHES measured at tunable illumination intensity and photon energy and characterize p-$Cd_{1-x}Mn_xTe$ single crystal revealing NDPC. Experimental results are analyzed with the Shockley-Read Hall model and the DLs structure is found. The suppressed electron signal is overcome by using a more powerful white laser source. We argue that such innovation strongly upgrades PHES capabilities, especially for semi-insulating semiconductors such as GaAs, InP, $HgI_2$, TlBr, $PbI_2$, SiC, CdTe-based materials, $CH_3NH_3PbCl_3$($Br_3$ or $I_3$) perovskites and also for organic semiconductors where the properties of both majority and minority carriers may be determined. We demonstrate the possibility to disclose both minority and majority carrier mobilities using the Hall mobility measured as a function of illumination intensity at certain photon energy $hv$ (obtained from PHES spectra) and intensity where hole or electron generation predominates. We show by photo-Hall effect measurements and Shockley-Read-Hall (SRH) charge kinetic simulations the reason of NDPC and ANPC appearance as well as the dependence of these effects on DL parameters such as carrier capture cross sections, photon capture cross sections, and DL concentration. We also demonstrate that NDPC can be a sign of bad transport properties as it may lead to the suppression of free carrier generation especially at a high-flux illumination.

## 2 Experimental

### 2.1 Material

To show principles of the PHES method, p-$Cd_{1-x}Mn_xTe$ sample was chosen. The single crystal was grown at the Chernivtsii National University by the vertical gradient freezing method with Mn content of 10%. The sample was cut from a wafer by diamond saw and polished on an alumina powder with 1 μm grain size. Afterwards, each side was chemo-mechanically polished for 30 s in 3% bromine-ethylene glycol solution [42,43]. Finally, bars adopted for the classical galvanomagnetic measurement were etched in 2% bromine-methanol solution for 1 min. After each chemical treatment the sample was rinsed in methanol and acetone.

Gold contacts for measurements were electrolessly deposited [44] on a pre-masked surface in 1% aqueous $AuCl_3$ solution. Resulting samples were rinsed in water and cleaned with acetone. The Fermi energy $E_F = E_v + 0.62$ eV found from the dark hole concentration is consistent with the given p-type resistivity of $3 \times 10^8$ Ωcm. The bandgap energy of the material is $E_g$ = 1.62 eV at the room temperature for given Mn content [38]. Laser-transient current technique showed no hole or electron transient signal in this sample.

## 2.2 Photo Hall effect measurements

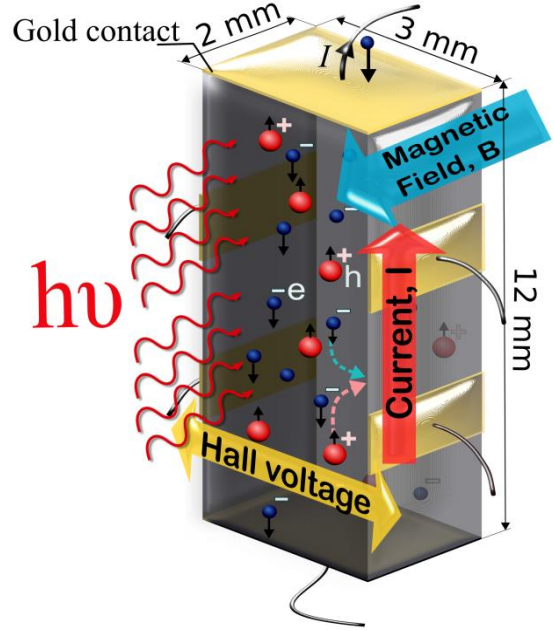

**Fig.1.** The basic principles of PHES method. The monochromatic illumination stimulates the free carrier generation. Photogenerated carriers are affected by Lorenz force and the electric field lengthwise the sample. The Hall voltage can be found under the steady-state conditions.

The principle of PHES is depicted in Fig.1. Bar-like p-type sample with dimensions of 2x3x12 mm³ was used in the classical six-contact Hall-bar shape convenient for galvanomagnetic measurements. Measurements were performed at room temperature with a constant magnetic field *B* of 1 T. The longitudinal voltage *V*, current *I*, and the transverse Hall voltage $V_H$ were measured directly from the experiment. The conductivity *Cond*, Hall coefficient $R_H$ and the Hall mobility $\mu_H$ were obtained by the following relations:

$$Cond = \frac{I \cdot L}{V \cdot S} \quad (1)$$

$$R_H = \frac{V_H d}{I \cdot B}, \quad (2)$$

$$\mu_H = Cond \cdot R_H = \frac{L}{S \cdot V} \cdot \frac{V_H d}{B}, \quad (3)$$

where *S, L,* and *d* are the cross section, distance between conductivity probes, and thickness of the sample correspondingly. Note that the obtained value of $\mu_H$ is determined by $V_H/V$ ratio. Therefore, $\mu_H$ remains unaffected by contingent transversal inhomogeneity of the sample induced for example by surface leakage current, if purely p-type character and spatially constant carrier mobility coextend through whole sample.

In our previous contribution [18] the moderate photon flux emitted by a conventional lamp source (halogen or mercury) filtered by a monochromator (the common output flux $< 10^{14}$ cm$^{-2}$s$^{-1}$) did not allow us sufficient minority carrier generation in the energy region below the Urbach tail. To overcome this obstacle we upgraded the equipment and used a more powerful white laser with a maximal filtered output photon flux of $1.3 \times 10^{15}$ cm$^{-2}$s$^{-1}$ here. The laser source provides nearly constant photon flux in the 0.6 -1.4 eV energy region [45] which is in the strongest interest for DLs spectroscopy. The sharp rise of the laser intensity near 1090 nm (25 nm segment) [45] was skipped by monochromator. The spectral resolution of the monochromator was measured using Ocean Optics spectrometer. The resolution (defined by FWHM) grows nearly linearly from 0.01 eV at 0.5 eV up to 0.11 eV at 1.3 eV. Linear interpolation of these values was used to correctly determine the DL threshold energies at intermediate energies.

The decrease of conductivity was not observed on other semi-insulating samples without NDPC in a similar energy region. To track the evolution of the photo-Hall signal under illumination intensity we measured spectra at three different constant photon fluxes. For the study of carrier properties at higher illumination flux we also used 0.95 eV and 1.46 eV laser diodes with maximal photon flux of $1.2 \times 10^{16}$ and $3 \times 10^{17}$ cm$^{-2}$s$^{-1}$, respectively.

## 3 Theory

### 3.1 The bipolar conduction and the interpretation concept of photo-Hall effect data

Overall in the paper we assume homogeneous sample with space independent carrier density and mobility both in the dark and under illumination. The transverse Hall voltage and longitudinal conductivity in the dark in p-type material depend on the dark minority ($n_0$) and majority ($p_0$) carrier concentrations and their mobilities. Dark carrier concentrations are further adjusted by photogenerated holes $p = p_0 + \Delta p$ and electrons $n = n_0 + \Delta n$ after illumination of the sample. It is important to note that $\Delta n \neq \Delta p$ and corresponding values mainly depend on filling of DLs in the band gap, illumination photon energy and intensity. In case of ANPC $\Delta n$ or $\Delta p$ may be even negative. The bipolar conductivity *Cond*, Hall coefficient $R_H$ photoconductivity *PhC* and Hall mobility $\mu_H$ are related to the free electron *n* and hole *p* concentrations by the relations:

$$Cond = q_e(\mu_h \cdot p + \mu_e \cdot n) \tag{4}$$

$$R_H = \frac{\mu_h^2 \cdot p - \mu_e^2 \cdot n}{q_e(\mu_h \cdot p + \mu_e \cdot n)^2} \tag{5}$$

$$PhC = q_e(\mu_h \cdot \Delta p + \mu_e \cdot \Delta n) \tag{6}$$

$$\mu_H = \frac{|\mu_h^2 \cdot p - \mu_e^2 \cdot n|}{\mu_h \cdot p + \mu_e \cdot n} \tag{7}$$

where $q_e$, $\mu_e$ and $\mu_h$ are an elementary charge, electron and hole mobility, respectively. Whenever the illumination photon energy ($h\nu$) reaches the trap energy counted relative to respective band, free electrons or holes are generated. This leads to a change of the spectral envelope and a rise of conductivity. In case of p-CdMnTe the response of $\mu_H$ to the photogenerated holes manifests in the fixation of the p-type character of the material often accompanied by an increase of $\mu_H$, see Eq. (7). This $\mu_H$ enhancement is interpreted as a result of screening of charge impurity scattering and smoothing of potential fluctuations. On the other hand, photogenerated electrons provide decrease of $\mu_H$ until the sign reversal of $V_H$ arises at $n = \left(\frac{\mu_h}{\mu_e}\right)^2 \cdot p$, which may be well approximated by $n \approx 0.01 \cdot p$ for CdTe-based compound materials. It is important to note here that whilst positive $V_H$ proves the p-type character of the material, due to the higher electron mobility the negative $V_H$ does not directly testify to the n-type. The comparison of $\mu_e$ and $\mu_H$ must be than included to solve this task.

Analyzing $\mu_H$ as a function of photon energy and light intensity enables us to determine the position of the DLs inside the band gap. Due to the higher electron mobility even smaller rise of electron concentration caused by secondary electron emission from the DL with threshold energy $E_t > E_F$ outlined by DL model No. 2 in Fig. 2 might produce $\mu_H$ decrease. The secondary emission involves thermal detrapping and secondary photon induced transition ($E_g - h\nu$) to the conduction band. A smart treatment must be chosen for the resolution of this task. The solution of the problem stems from different mechanisms of free electron generation. The photon induced transition from the filled DL below $E_F$ to the conduction band satisfies the rule $\Delta n = \Delta n_t^{empt} + \Delta p$, where $\Delta n_t^{empt}$ is a concentration of photo induced empty states of the deep level (see Fig. 2). The secondary electron emission produced by empty (or nearly empty) DL with the energy $E_t>E_F$ is ruled by a similar relation $\Delta n + \Delta n_t = \Delta p$, where $\Delta n_t$ is a concentration of trapped electrons on the deep level. As we can see, this alternative transition is followed by predominant hole generation and the photogenerated hole concentration must be higher $\Delta p > \Delta n$ than the concentration of photogenerated electrons. The distinctions between the electron excitation mechanisms from DL localized below or above $E_F$ were used for determining DL position in the band gap using the dual wavelength excitation [17]. A similar concept can be used in a p-type material where $\mu_h > \mu_e$.

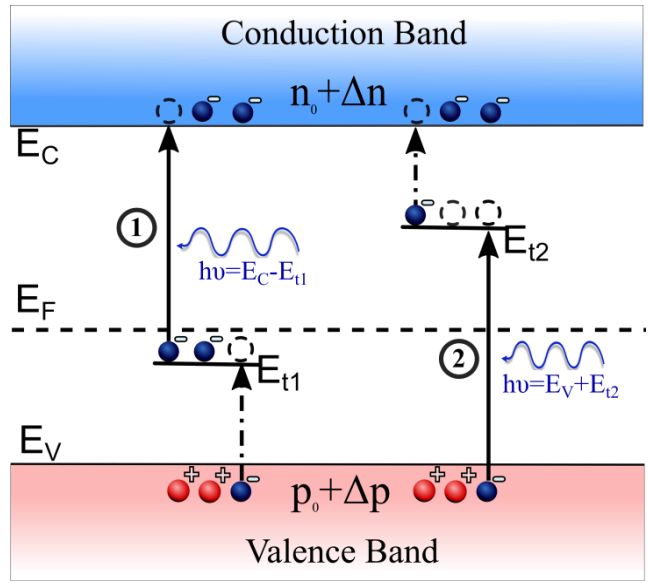

**Fig. 2.** Competing deep level models No 1 and No 2 . Solid and dashed arrows show direct and secondary carrier generation processes correspondingly. $E_F$ show Fermi level position in p-type material.

## 3.2 Shockley-Reed-Hall charge kinetic simulations.

The Shockley-Read-Hall (SRH) theory [46] complemented by illumination-mediated deep level - band transitions shown in Fig.3 is used to analyze in detail two alternative DL models and NDPC appearance. The evaluation of corresponding charge dynamics is performed in steady state illumination regime. Equations (8)-(10) represent the electron, hole and DLs occupancy dynamics.

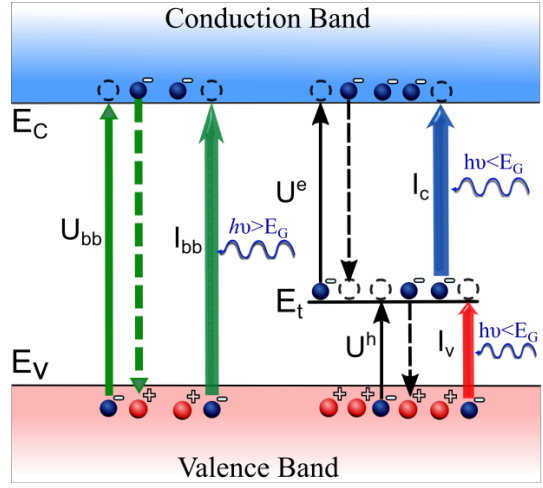

**Fig. 3.** Shockley-Reed-Hall carrier generation-recombination model. Antiparallel solid and dashed arrows show thermal carrier generation and recombination processes (band to band or DL-assisted). Solid upward arrows correspond to photon assisted free carriers generation (band to band or DL-assisted).

$$\frac{\partial n}{\partial t} = -\sum_i U_i^e + \sum_i I_{ci} + I_{bb} - U_{bb}, \qquad (8)$$

$$\frac{\partial p}{\partial t} = -\sum_i U_i^h + \sum_i I_{vi} + I_{bb} - U_{bb}, \qquad (9)$$

$$\frac{\partial n_{ti}}{\partial t} = U_i^e - I_{ci} - U_i^h + I_{vi}, \qquad (10)$$

$$U_i^e = \sigma_{ei} v_e [(N_{ti} - n_{ti})n - n_{ti}n_{1i}], \qquad (11)$$

$$U_i^h = \sigma_{hi} v_h [n_{ti}p - (N_{ti} - n_{ti})p_{1i}], \qquad (12)$$

$$I_{ci} = I\tilde{\alpha}_{ei}n_{ti}, \qquad (13)$$

$$I_{vi} = I\tilde{\alpha}_{hi}(N_{ti} - n_{ti}), \qquad (14)$$

$$I_{bb} = G_{bb}I, \qquad (15)$$

$$U_{bb} = C_{bb}(np - n_i^2), \qquad (16)$$

Here $n$, $p$, $n_{ti}$, $U_i^e$ ($U_i^h$), $I_{bb}$ and $U_{bb}$ are the densities of free electrons, free holes, electrons trapped in the $i$-th level, electron (hole) net recombination rate at the $i$-th level, the interband light induced generation rate, and the band-to-band net recombination rate. The quantities defining recombination rates $n_i$, $N_{ti}$, $\sigma_{ei}$, $\sigma_{hi}$, $v_e$ and $v_h$ in Eqs. (10-14) are intrinsic carrier density, $i$-th DL density, electron and hole thermal capture cross sections, and electron (hole) thermal velocities, respectively. Symbols $n_{1i}$ and $p_{1i}$ stand for electron and hole densities in case of Fermi level $E_F$ being set equal to the DL ionization energy $E_{ti}$ [46]. The effect of illumination on the $i$-th DL occupancy is defined by $I_{ci}$ ($I_{vi}$) generation rate from $i$-th level to the conduction (valence) band where $I$, $\tilde{\alpha}_{ei}$, and $\tilde{\alpha}_{hi}$ are the photon flux and photon capture cross sections relevant to the conduction and valence band transition respectively. The interband light induced generation rate $I_{bb}$ can be neglected for $h\nu<E_g$.. The band-to-band recombination constant $C_{bb} = 10^{-11}$ cm$^2$ can be estimated for Cd$_{1-x}$Mn$_x$Te according to Ref [47]. The band-to-band recombination rate is typically much less in comparison with DL-assisted recombination rate and it can be neglected. The values of photon capture cross section were chosen to fit the shape of PHES spectra and the order of magnitude was chosen according to Ref. [17]. Both photon energy and photon flux dependencies were fitted simultaneously by the same parameters. The charge neutrality was assured by the neutrality equation

$$p - n - \sum_i n_{ti} = p_0 - n_0 - \sum_i n_{t0i}, \qquad (17)$$

where the dark densities $p_0$, $n_0$, and $n_{t0i}$ on the right-hand side are defined by the position of Fermi energy $E_F$. The solution of equations (8)-(10) and (17) significantly simplifies in the steady state regime, where the time derivatives are set to zero. The electron and hole lifetimes $\tau_e$ and $\tau_h$ can be found by relations:

$$\tau_e = \frac{1}{\sum_i v_{ei}\sigma_{ei}(N_{ti}-n_{ti})}, \qquad (18)$$

$$\tau_h = \frac{1}{\sum_i v_{hi}\sigma_{hi}n_{ti}}. \qquad (19)$$

The occupation of DLs can be conveniently described by a dimensionless filling factor (FF) which is defined as the ratio of DL electron occupation and the total DL concentration $f_i = n_{ti}/N_{ti}$. The change of the DL occupation from the equilibrium under illumination intensity $I$ can be presented by reduced filling factor $F_i(I)$, $-1 < F_i < 1$:

$$F_i(I) = f_i(I) - f_i(0) \qquad (20)$$

The occupation and depletion of the deep level can be tracked by the positive or negative sign correspondingly.

# 4 Results and Discussion

## 4.1 Deep levels, carrier mobility, and negative differential photoconductivity

Photoconductivity and Hall mobility spectra measured in the energy range 0.6-1.8 eV under three different photon fluxes are presented in Fig. 4. One can see rises and the change of the PhC slope at 0.63, 0.9, 1.0 and 1.3 eV photon energies related to the presence of DLs near these threshold energies. It is important to note here that PhC starts to saturate and change the sign of the derivative in the energy region 0.96 -1.3 eV, Fig. 4 (a). This effect is known as the negative differential photoconductivity. The Hall mobility $\mu_H$ starts to increase from 0.62 eV up to 0.96 eV where it starts to saturate, Fig. 4 (b) Region I. The increase of $\mu_H$ in the p-type sample is reached by the free hole generation according to Eq. (7) associated with the transitions to DLs from the valence band in this energy region as mentioned in section 3.1. The $\mu_H$ decrease can be observed at $h\nu > 0.96$ eV, Region II, and the negative sign of $V_H$ is reached near 1.3 eV for a maximal photon flux of $1.3 \times 10^{15}$ cm$^{-2}$s$^{-1}$. The decrease of $\mu_H$ is caused by free electrons generated in this region. One can see that the sign change of $\mu_H$ depends on the illumination intensity being shifted to higher photon energy at decreasing photon flux $9 \times 10^{14}$ cm$^{-2}$s$^{-1}$ and $4 \times 10^{14}$ cm$^{-2}$s$^{-1}$. In contrast to the abrupt change of $\mu_H$ induced by photo-electrons, *PhC* remains stable in the energy region 1.0-1.4 eV, which proves the stable hole density dominating in *Cond* and *PhC*. Once a negative sign of $V_H$ is reached, $\mu_H$ increases with the further increase of the photon energy in agreement with Eq. (7).

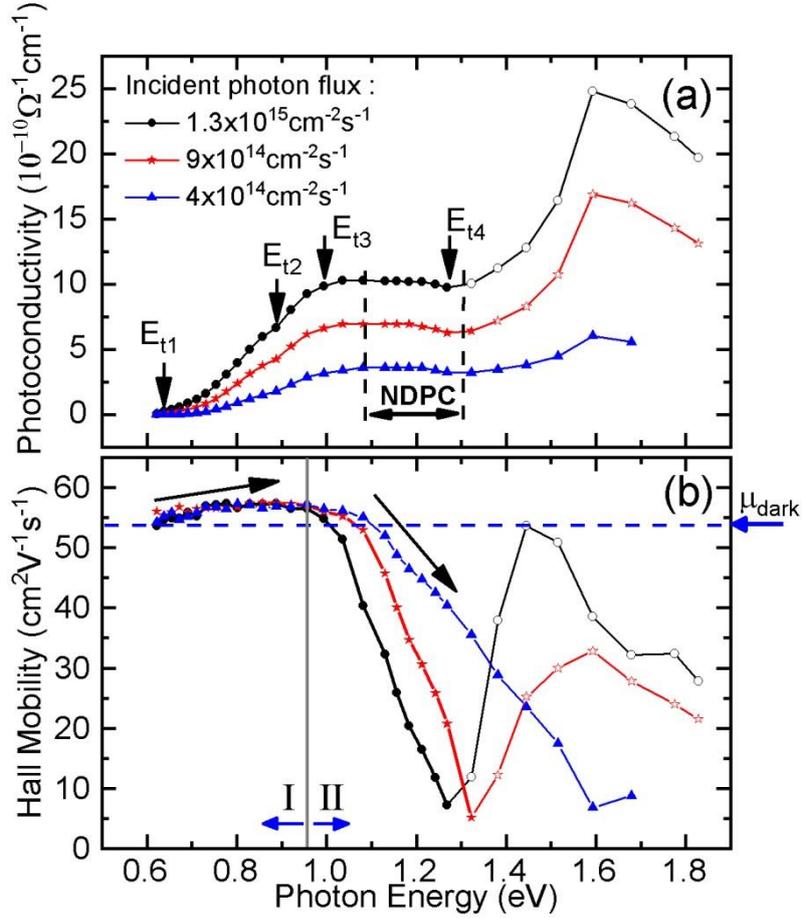

**Fig.4.** Photoconductivity (a) and Hall mobility (b) spectra at different photon fluxes. Vertical arrows show the deep level threshold energy. Here and henceforth the points with the negative $V_H$ sign are shown by empty symbols.

To find the position of DL relative to $E_c$ and $E_v$, i.e. to distinguish the level model 1 and 2 outlined in Fig. 2, free carrier concentrations spectra are needed. For evaluation of $n$ and $p$ from Eqs. (6), (7) in the mixed conductivity regime corresponding mobilities $\mu_e$ and $\mu_h$ must be known. The theoretically predicted values were often chosen in the photo-Hall measurements in CdTe [6] and perovskites [12]. However, such choice is not convenient for this analysis and it can lead to incorrect conclusions.

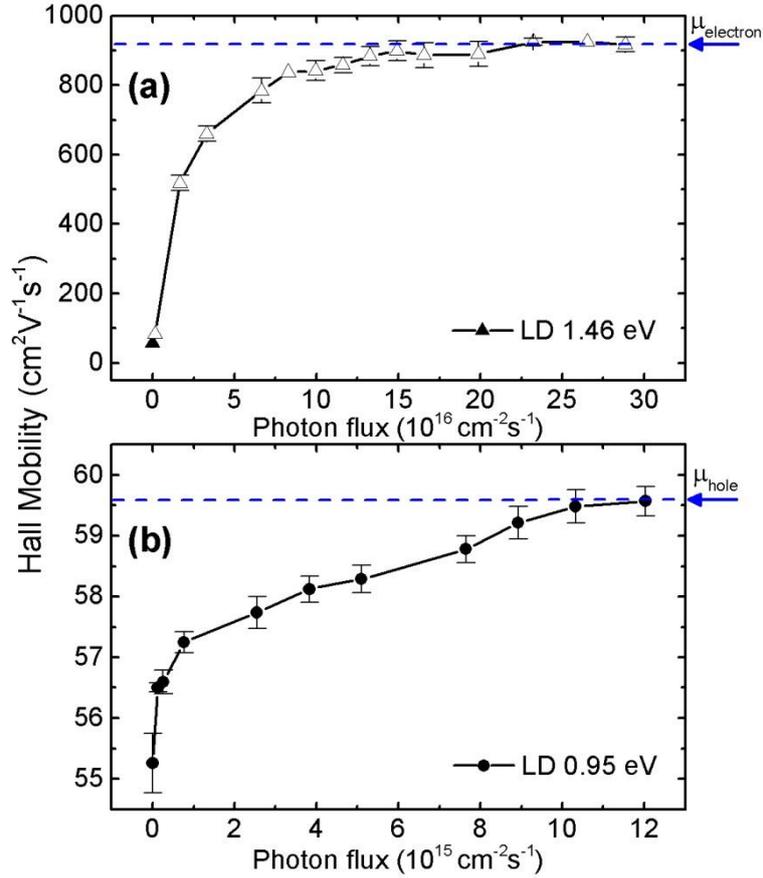

**Fig.5.** Hall mobility measurements at extended photon fluxes with photon energies 1.46 eV (a) and 0.95 eV (b). Horizontal arrows and dashed lines show saturation values of the Hall mobility associated with electron and hole mobility. The solid and empty symbols represent data with the positive and negative sign of $V_H$.

The real value of mobility affected by actual free carrier density must be measured directly. According to Eq. (7) electron and hole mobilities can be found in the conditions where only one type of carriers is dominating. As we already found, energy regions I and II, where photoexcited holes (with maximal hole mobility at ~0.96 eV), and electrons (with maximal electron effect at ~1.45 eV) dominate; $\mu_H$ was measured as a function of photon flux at these energies using laser diodes as an enhanced illumination source. We show in Fig. 5 the rapid increase of $\mu_H$ both for electrons in Fig.

5(a), where $\mu_H$ saturates at 925±11 cm$^2$s$^{-1}$V$^{-1}$, and for holes in Fig. 5(b) with saturation at 59.6±0.4 cm$^2$s$^{-1}$V$^{-1}$. In this paper, saturated values of the Hall mobility observed in Fig. 5 (a)-(b) are further associated with free carriers mobility.

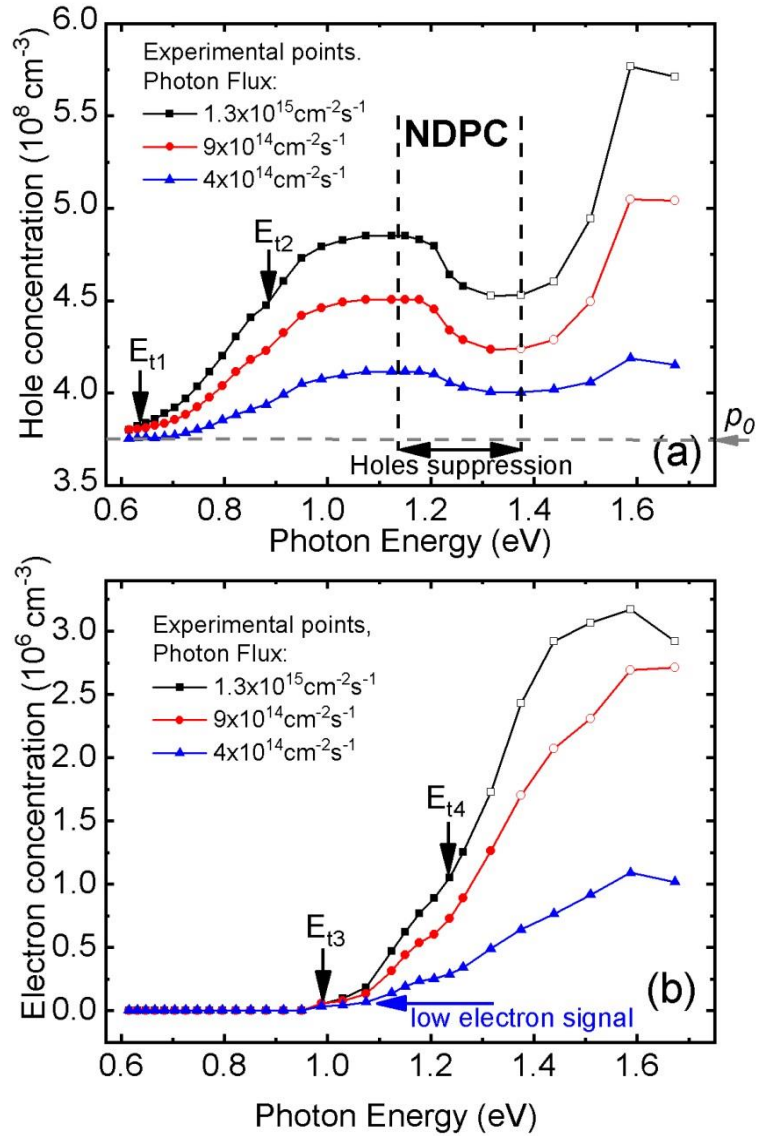

**Fig. 6.** Hole (a) and electron (b) concentration spectra at different photon fluxes. Vertical arrows show DL threshold energies. Vertical dash lines separate the region of negative differential photoconductivity.

Free electron and hole concentration spectra in the energy range 0.6-1.8 eV calculated from Eq. (6-7) and data from Fig. 4 using saturation values of the Hall mobility are presented in Fig.6. One can see the rise of hole concentration at the energies 0.63 eV and 0.9 eV. These changes of the spectra are associated with a photon induced transition from the valence band to DLs with threshold energies $E_{t1} = E_V + 0.63$ eV, and $E_{t2} = E_V + 0.9$ eV. As can be seen from Fig. 6(b), no free electron generation is detected in the energy region $h\nu < 0.96$ eV. Therefore, free electrons are neglected in region I. Corresponding transitions and DLs threshold energies are schematically shown in Fig. 7. The rise of the electron concentration can be observed at 0.96 eV and 1.25 eV photon energies and it is associated with electron transitions from the DLs localized below or close to Fermi level with threshold energies $E_{t3} = E_c - 1.0$ eV and $E_{t4} = E_c - 1.3$ eV (taking into account the resolution of the output light), which are the subject of model 1 discussed further, or secondary transitions from alternative DLs $E_{t3} = E_V + 1.0$ eV and $E_{t4} = E_V + 1.3$ eV related to model 2. The rise of the electron concentration $n$ is followed by saturation and decrease of the hole concentration $p$. In view of the fact that NDPC appears at similar photon energy where electrons start to operate, we deduce that principal recombination level responsible for electron-hole recombination rules the process. The *PhC* decrease in this region is damped by the free-electron rise in NDPC region. It is important to note that DLs threshold energies can be mistakenly shifted at lower illumination photon fluxes and such tendency can be seen in Fig. 6(b) near 1.0 eV photon energy. The conventional light sources [18] often used with monochromator have maximal output photon flux of $10^{14}$ cm$^{-2}$s$^{-1}$. The electron concentration at photon fluxes less than $4\times10^{14}$ cm$^{-2}$s$^{-1}$ is nearly undetectable.

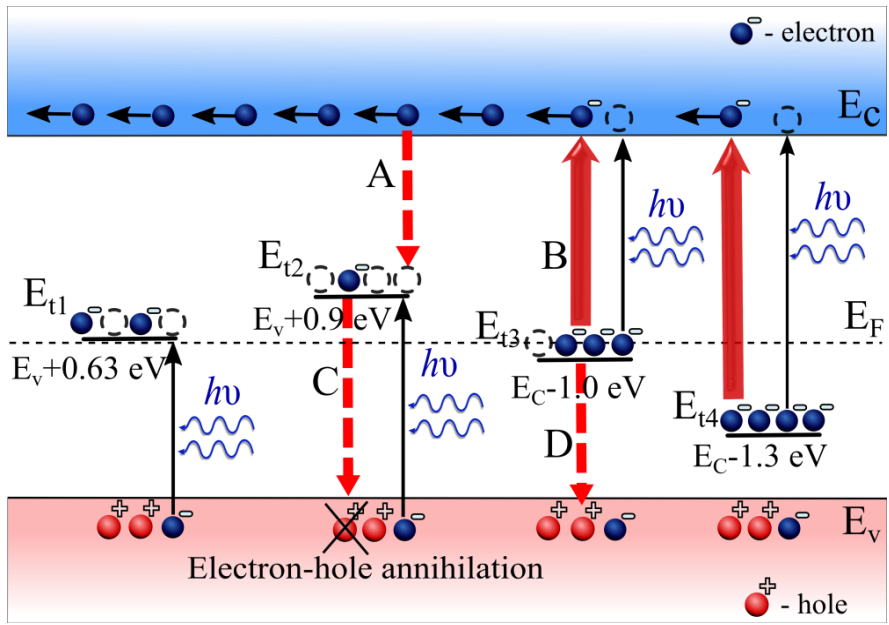

**Fig.7.** The defect model 1 with pondered transitions of the deep levels. Full upward black arrows delineate principal optical excitation. Bold and dashed red arrows show the generation-recombination traffic which leads to fast electron-hole annihilation and decrease (saturation) of hole concentration under illumination.

## 4.2 Deep level detection and charge carrier Shockley-Reed-Hall simulations

In this section we introduce SRH simulation results which represents the fit of all experimental data according to Eqs. (8)-(17) with parameters from Table 1, denoted as model 1(a). Models 1(b) and 1(c) represent simulations where the variation of DL parameters takes place to see the influence of DL properties on the studied effects, see Table 2. Model 2 is a simulation with alternative DLs position above $E_F$ mentioned in Section 4.1 above.

**Table 1.** Deep level parameters of model 1(a).

| DL | $E_{t1}$ | $E_{t2}$ | $E_{t3}$ | $E_{t4}$ |
|---|---|---|---|---|
| Position in the band gap, eV | $E_V + 0.63$ | $E_V + 0.9$ | $E_C - 1.0$ | $E_C - 1.3$ |
| $N_t$, cm$^{-3}$ | $10^{13}$ | $6.5 \times 10^{12}$ | $3.5 \times 10^{12}$ | $8 \times 10^{12}$ |
| $\sigma_e$, cm$^2$ | $3 \times 10^{-17}$ | $9 \times 10^{-13}$ | $5 \times 10^{-14}$ | $2 \times 10^{-14}$ |
| $\sigma_h$, cm$^2$ | $4 \times 10^{-16}$ | $5 \times 10^{-14}$ | $5 \times 10^{-13}$ | $3 \times 10^{-19}$ |
| $n_{ti}/N_{ti}$* | 0.35 | $10^{-5}$ | 0.49 | 1 |

*The relative deep level occupation at equilibrium.

Concentrations of holes and electrons calculated according to Shockley-Reed-Hall model simulations completed by four DLs from model 1 and model 2 are shown in Fig. 8 by dash and dash-dot lines, respectively. DL threshold energies $E_{t1}$ and $E_{t2}$ are fixed in both models as these energies appear safe. The variation of the model parameters over a wide range according to the DL parameters in the literature [48,49] showed that competing model 2 fails to provide electron concentration $\Delta n > \Delta p$ in the region $hv > 1.0$ eV, because an emission like electron generation is followed by strong hole generation. Moreover, $p$ decrease and NDPC are not reached for the same reason. The behavior of free holes and electrons in this region excludes the possibility of another DL localization in the band gap (with $E_t > E_F$), where electron rise is produced by secondary electron generation. On the other hand, model 1 shows good results in simulation of the PHES spectra, see curve model 1(a) in Fig 8. This proves the validity of the proposed PHES deep level detection concept. Note that DLs $E_{t1}$, $E_{t2}$ and $E_{t3}$ have similar activation energies which can be hardly resolved by photoluminescence or thermal activation methods where DL band overlapping occurs [16,18]. According to model 1(a) and Table 1 the electron lifetime $\tau_e = 5 \times 10^{-9}$ s is controlled by electron trap DL $E_{t2}$ due to strong recombination. DL $E_{t3}$ plays a minor role in the electron trapping due to lower concentration and electron capture cross section. The majority hole lifetime $\tau_h = 10^{-7}$ s is controlled mainly by the hole trap $E_{t3}$.

**Table 2.** Variation of deep level parameters of model 1(a).

| Model No: | Changed parameters: |
| --- | --- |
| model 1 (b) | $\sigma_{h2,\,1(b)} = 10 \times \sigma_{h2} = 5\times10^{-13}$ cm$^2$ |
| model 1 (c) | $N_{t2,\,1(c)} = 10 \times N_{t2,\,1(c)} = 6.5\times10^{13}$ cm$^{-3}$ |
| model 2 | $E_{t3,\,(2)} = E_V + 1.0$ eV; $E_{t4,\,(2)} = E_V + 1.3$ eV |

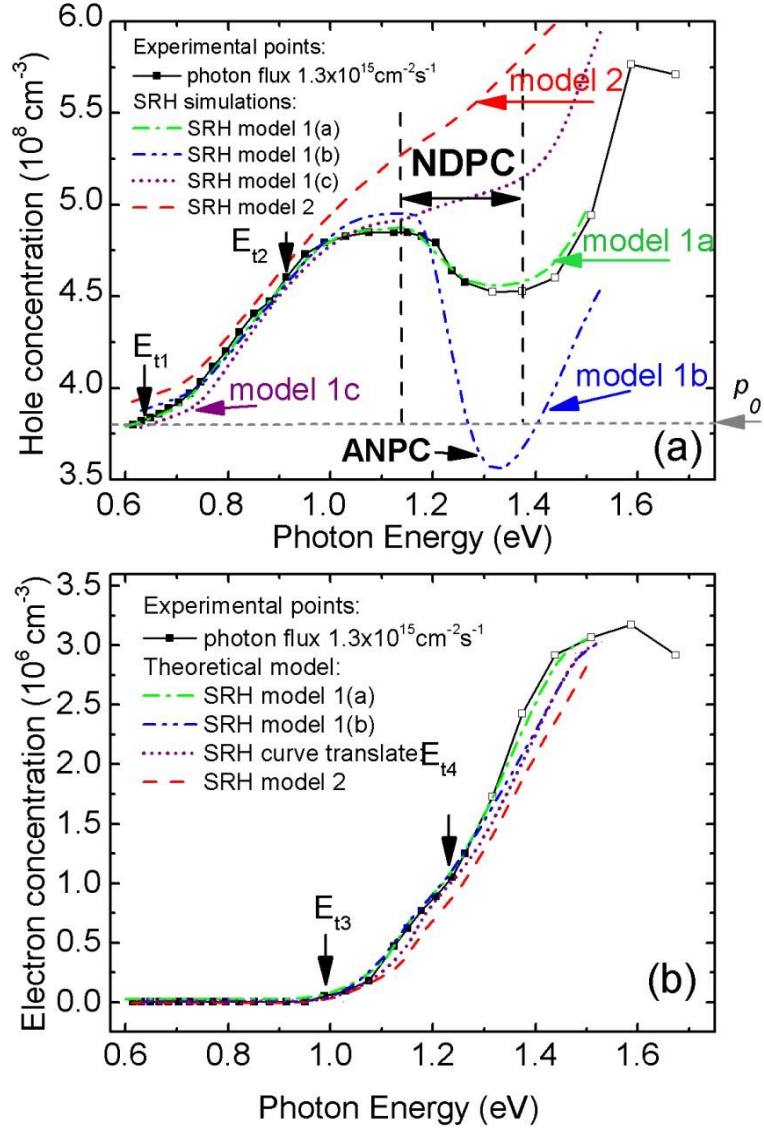

**Fig. 8.** The dashed (dash-dot) curves show SRH model simulations for two possible models (model 1(a) and model 2). Models 1(b) and 1(c) show SRH simulation in case of increased capture cross section and defect concentration of the recombination DL, respectively. Theoretical curves are calculated for the maximum photon flux and linearly shifted upwards/downwards with an increment of 1% for better visualization. The solid line with squares shows experimental data from Fig. 6 for the maximum photon flux.

## 4.3 Negative differential photoconductivity characterization

According to experimental results and SRH simulations, the electron generation in the region II is provided by DLs $E_{t3}$ and $E_{t4}$ localized below $E_F$ (model 1) and the hole concentration rises are induced by $E_{t1}$, and $E_{t2}$. According to model 1(a) the effect of NDPC detected in Fig. 4 (a) is produced by the decrease of the hole concentration confirmed in the experiment, Fig. 6 (a). The decrease of the majority carrier's concentration under sub-bandgap illumination is reached by minority carrier's fast recombination supported mainly by DL $E_{t2}$ and as a result of electron-hole annihilation, as shown in Fig. 7 by bold and dashed red arrows (processes A-D). DL $E_{t2}$ is empty enough to provide a strong recombination channel for free electrons. The decrease of hole and increase of electron concentrations match the experimental points in NDPC region and confirm the reliability of the proposed deep level assisted NDPC model.

For further calculation, we chose trap $E_{t2}$ as the most important recombination channel to show the dependence of the main NDPC effects on DL parameters. Another DLs are occupied and have insufficiently low carrier capture cross sections. Therefore they can hardly serve as a strong recombination channel. The effect of hole concentration decrease associated with NDPC depends on the capture cross section $\sigma_{h2}$ which represents the hole recombination rate of the DL, see Eq.(12). The curve of model 1(a), in Fig. 8 shows results for $\sigma_{h2} = 5 \times 10^{-14}$ cm$^2$ and the model 1(b) represents simulation where $\sigma_{h2,\ 1(b)} = 5 \times 10^{-13}$ cm$^2$. The hole concentration decrease can be stimulated even more expressively and the concentration $p < p_0$ can be reached by the $\sigma_{h2}$ increase. A similar effect can be produced by electron photo-generation ($\sim I\tilde{\alpha}_{ei}$ in Eq. (13)) increase from the DLs $E_{t1}$, $E_{t3}$, or $E_{t4}$. The effect of the carrier decrease under illumination below the dark equilibrium values is known as absolute negative photoconductivity. We found that the decrease in the hole concentration in NDPC region is not observed for capture cross sections $\sigma_{h2} < 2 \times 10^{-14}$ cm$^2$.

To show the effect of the recombination DL concentration on $p$ decrease we developed model 1(c) where the concentration of the recombination channel DL $E_{t2}$ ($2 \times 10^{13}$ cm$^{-3}$) is higher by an order of magnitude than in model 1(a) ($6.5 \times 10^{12}$ cm$^{-3}$), see spectra in Fig. 8 (dotted purple line). Despite the fact that electron lifetime is dropped from the value $5 \times 10^{-9}$ s to $5 \times 10^{-10}$ s the decrease of $p$ in the NDPC region is completely suppressed. The main reason of this result is comprised in the increase of DL $E_{t2}$ concentration, which also implies an increase of the terms $\sigma_{hi}v_h(N_{ti} - n_{ti})p_{1i}$ and $I\tilde{\alpha}_{hi}(N_{ti} - n_{ti})$ responsible for thermal and optical generation of free holes, see Eqs. (9), (12) and (14). Meanwhile, the recombination term $\sigma_{hi}v_h n_{ti}p$ and electron generation term $I_{ci} = I\tilde{\alpha}_{ei}n_{ti}$ are increasing slowly due to emptiness of the DL above $E_F$. That's why free hole concentration increases in NDPC region and does not decrease when DL concentration increases. It is important to remind that the deep level $E_{t3}$ acts as an alternative recombination channel in case of the absence of the deep level $E_{t2}$. Therefore, to define the lower concentration of the recombination channel at which NDPC is observed, DL $E_{t3}$ must be included for analysis. We have found that the decrease of the hole concentration under illumination was observed when DL concentrations met the following relation:

$$10^{10} cm^{-3} < \frac{N_{t2}+N_{t3}}{2} < 2 \times 10^{13} cm^{-3}, \tag{21}$$

As one can see, even in the material with relatively low defect concentration [5] NDPC can be stimulated. Multiplying the left part on the limit value for capture cross section we get NDPC lower limit rule for CdMnTe material.

$$N_t \sigma_h > 2 \times 10^{-4} cm^{-1}, \tag{22}$$

It should be noted that in case of absence of the electron generation channels (DLs $E_{t3}$ and $E_{t4}$ localized below $E_F$) the NDPC effect vanishes, see Fig. 8(a) model 2. This illustrates an important fact that both recombination and minority carrier generation channels must be present in the material to observe negative differential photoconductivity effect.

## 4.4 Charge carrier transport simulations under high illumination fluxes

The band gap illumination is often used to study device performance instead of X- and gamma-ray or alpha particle radiation sources [50,51]. The carrier concentrations of $Cd_{1-x}Mn_xTe$ and detector grade CdTe (CdTe sample without NDPC effect) were studied under extensive 1.46 eV laser illuminations, to show the influence of discovered DLs and NDPC on the material absorption properties. One can see in Fig. 9 that the hole concentration starts to saturate and reaches the saturation value of $1.8 \times 10^9$ cm$^{-3}$ at photon flux $7 \times 10^{16}$ s$^{-1}$cm$^{-2}$ in $Cd_{1-x}Mn_xTe$ sample. The further increase of the photon flux does not enhance the free hole concentration. The suppression of *p* is followed by a slow superlinear increase of *n*. The proposed NDPC deep level model 1(a) presented by a solid curve is in a good agreement with experimental data. Experimental data and SHR simulations with DLs positions in the band gap discovered by PHES show that such hole suppression and superlinear electron raise are the consequence of the recombination through half empty DL $E_{t2}$ (and partially DL $E_{t3}$ ) and further electron-hole annihilation. Numerical simulations also show that in the absence of the electron generation channel according to model 2 no NDPC can be observed. In case of enhanced recombination channel, model 1(b), the appearance of NDPC effect is followed by ANPC at higher fluxes as it was predicted by the PHES spectra simulation, Fig. 8 (a) NDPC region.

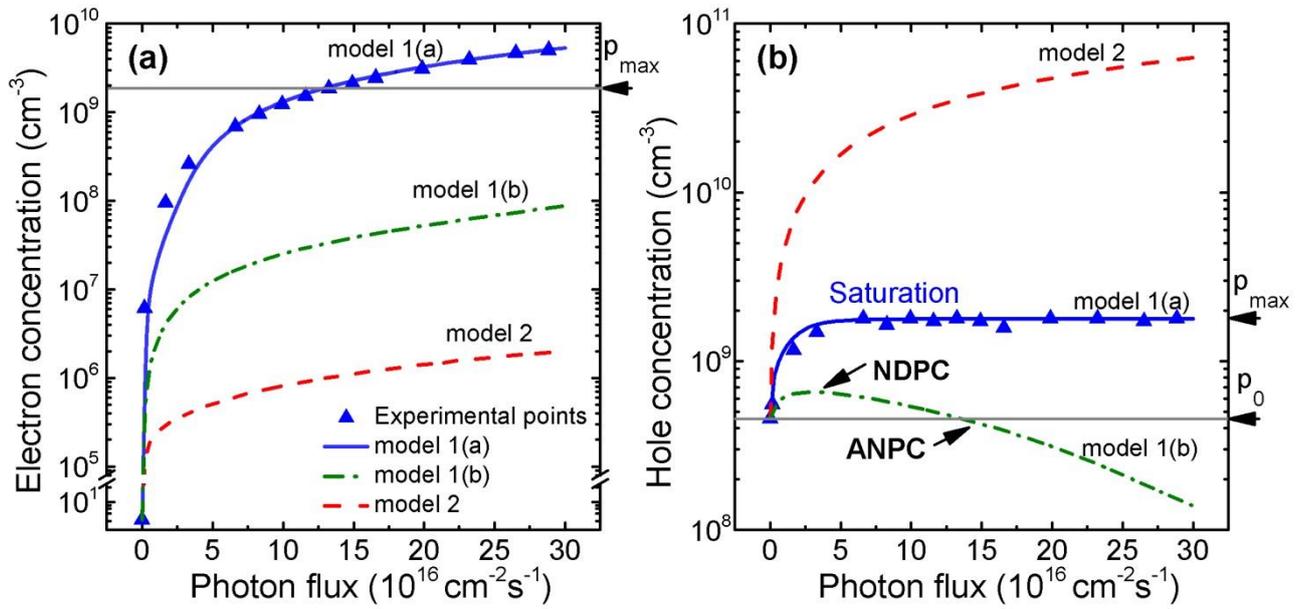

**Fig. 9.** (a) Electron and (b) hole concentration as a function of photon flux at 1.46 eV high-flux illumination in $Cd_{1-x}Mn_xTe$. The dashed curves show SRH simulations with DLs from Fig. 7 (model a).

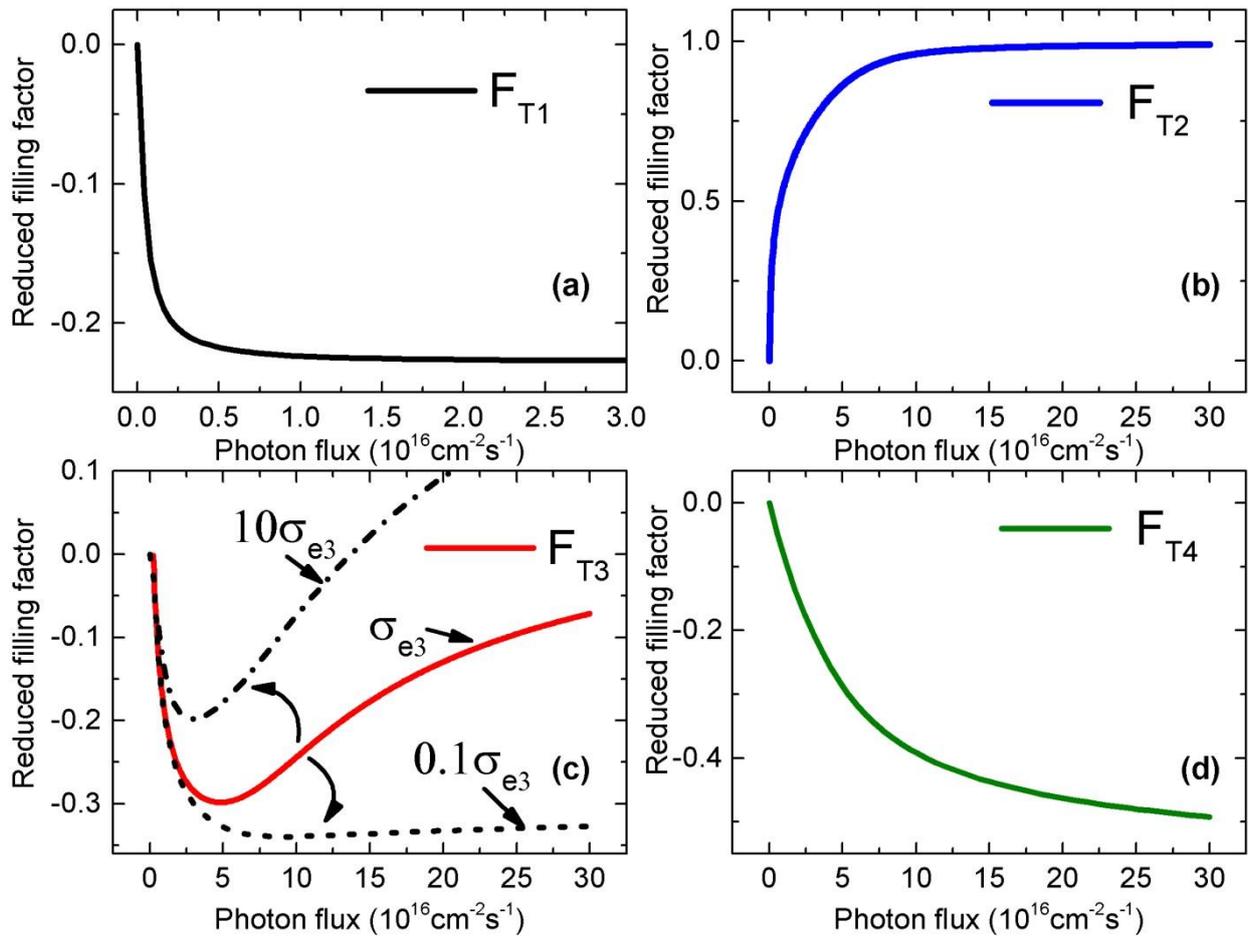

**Fig. 10.** Deep level reduced filling factor calculated by SRH theory for model 1(a) with deep levels from Fig. 7.

The occupancy of deep levels of model 1(a) as a function of photon flux is shown in Fig. 10 by means of reduced filling factor $F_i(I)$. The electron generation at 1.46 eV photon energy is supported by DLs $E_{t1}$, $E_{t3}$ and $E_{t4}$, as can be seen from the depletion of the corresponding DLs. The recombination of minority electrons is driven by DLs $E_{t2}$ and $E_{t3}$ where the increased occupation of DLs is observed. Note the nontrivial occupation shape of DL $E_{t3}$. According to Eq. (8) the generation rate $I_{ci}$ is connected to recombination rate $U_i^e$. With increasing illumination intensity both terms are increasing, see Eqs. (8) and (13). Due to initial depletion tendency of DL $E_{t3}$, the electron thermal emission part of the net recombination rate $\sigma_{ei} v_e n_{ti} n_{1i}$ is decreasing in comparison to equilibrium state. On the contrary, the electron capture rate $\sigma_{ei} v_e (N_{ti} - n_{ti}) n$ is increasing and predominates the thermal emission part. Similar tendencies are observed for the hole associated processes $U_i^h$ and $I_{vi}$. The steady state conditions for trapped electrons (holes) are preserved in Eq.10. Note that the electron generation rate $I_{ci}$ is proportional to $n_{ti}$ and hole generation rate $I_{vi}$ is proportional to $N_{ti} - n_{ti}$. Due to the initial tendency of electron depletion of partly filled DL, hole generation rate is increasing faster than the electron one. This synergy of facts leads to nontrivial DL $E_{t3}$ occupation. To emphasize the influence of the electron recombination trough DL $E_{T3}$ on non-trivial occupation behavior (process A, Fig. 7), we present the occupancy profiles for enlarged and reduced electron capture cross section $\sigma_{ei}$ in Fig. 10 (c). The understanding of complicated dynamics of the space charge in radiation detectors utilized at high flux has so far been a serious problem [40,52–55].

# 6 Conclusion

Photo-Hall effect spectroscopy with enhanced illumination has proven as a powerful tool for material characterization and deep levels detection. The enhanced illumination allowed us a separate experimental investigation of both minority and majority carriers. Absolute positions of deep levels in the band gap hardly accessible by conventional spectroscopy methods were determined by PHES.

The data obtained from PHES were used for developing a charge generation-recombination model. We showed that such model is very useful for a detailed analysis and understanding of effects like negative differential photoconductivity, absolute negative photoconductivity, etc. These multiple benefits and the resolved problem of DL energy detection place photo-Hall effect measurements with enhanced monochromatic illumination over other material characterization methods.

We explained the appearance of negative differential photoconductivity by fast minority carrier recombination through the nearly empty DLs and subsequent electron-hole annihilation, which leads to a decrease or saturation of majority carrier concentration. We showed that both recombination and generation channels are necessary to observe NDPC. We also presented a comprehensive analysis of NDPC effect by means of SRH simulations. We showed the dependence of the effect on DL parameters such as carrier capture cross sections and DL concentration, as well as the connection between negative differential photoconductivity and absolute negative photoconductivity. The influence of such an effect on the material properties was studied in the regime of enhanced laser illumination at $hv$ = 1.46 eV. It was argued that the appearance of NDPC depreciates charge transient properties and leads to undesirable loss of photo-carriers. NDPC can be observed even in a material with relatively low defect concentration.